\documentclass[twocolumn, amsmath,superscriptaddress, amsfonts,prb]{revtex4}
\usepackage{graphicx}
\begin{document}

\title{Nucleation and growth of metal whiskers}
\author{V. G. Karpov}\email{victor.karpov@utoledo.edu}\affiliation{Department of Physics and Astronomy, University of Toledo, Toledo, OH 43606, USA}
\begin{abstract}
The existence of metal whiskers is attributed to the  energy gain due to electrostatic polarization of needle shaped metal filaments in the electric field induced by surface imperfections: contaminations, oxide states, etc. A proposed theory provides closed form expressions for the whisker nucleation and growth rates, explains the range of whisker parameters and effects of external biasing. It predicts a well controlled whisker growth on any metal surface via generating surface plasmon polariton excitations.

\end{abstract}

\date{\today}

\maketitle
\section{Introduction}\label{sec:intro}
\begin{figure}[b!]
\includegraphics[width=0.5\textwidth]{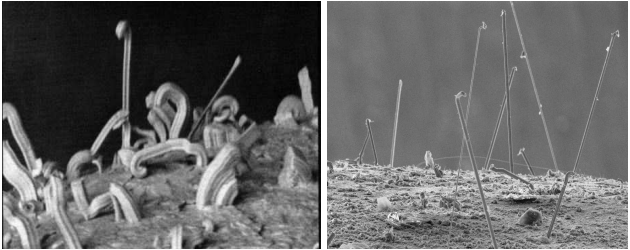}
\caption{SEM pictures of tin (left) and zinc (right) whiskers. Reproduced from the NASA photogallery. \cite{photo} \label{Fig:whiskimage}}
\end{figure}

Metal whiskers are hair-like protrusions grown at surfaces of some metals, tin and zinc representing important examples illustrated in Fig. \ref{Fig:whiskimage}. \cite{nasa1,calce,nasa2,bibl,nasa3,photo} In spite of being omnipresent and leading to multiple failure modes in electronic industry, the mechanism behind metal whiskers remains unknown after more than 60 years of research. Here, we present such a mechanism consistent with many published observations and providing verifiable predictions.

\bigskip

As a brief survey of relevant data, \cite{nasa1,calce,nasa2,bibl,nasa3,photo,shetty2004,crandall,munson,fang2006} we mention that whiskers grow up to $\sim 10$ mm in length and vary from $\sim 1$ nm to 30 $\mu$m in diameter; their parameters are characterized by broad statistical distributions. The metal surface conditions play a significant role making oxide structure and various contaminations important factors determining whisker concentration, growth rate and dimensions; however, the metal grain size appears to be of much less significance.  Various additives also can have significant effects, such as e. g. small concentration of Pb strongly suppressing tin whiskering. External electric bias exponentially increases whiskers growth rate. A comprehensive review of experimental data on the most studied tin whiskers before the year of 2003 was given in a monograph \onlinecite{galyon2003}.

Multiple attempts to understand the mechanisms of whiskers growth (see e.g. Refs. \onlinecite{nakai2009,sobiech2008,smetana2007,barsoum2004}) revolved around the role of surface stresses relived by whisker production, dislocation effects, and oxygen reactions.

The mechanism proposed here is qualitatively different as driven by the existence of strong electric field $E$ above the metal surface. The field is generally due to surface imperfections, such as oxide, ion contaminations, and interfacial states. The appearance of whiskers is described as the field induced nucleation. It is triggered by the energy gain $-{\bf p\cdot E}$ due to interaction between the field $E$ and its induced whisker dipole ${\bf p}=\alpha {\bf E}$ where $\alpha$ is the polarizability. The latter is anomalously strong for the needle shaped metallic particles that serve as whiskers' nuclei.

\section{Field induced nucleation of whiskers}
\begin{figure}[t!]
\includegraphics[width=0.35\textwidth]{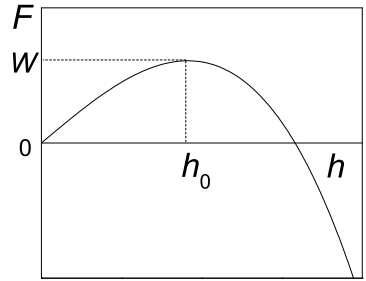}
\caption{Free energy of a whisker vs. its length. \label{Fig:F}}
\end{figure}

The electrostatic energy gain in the electric field can be represented as \cite{kaschiev2000,warshavsky1999,isard1977}
\begin{equation}W_E=-\varepsilon\alpha  E^2 \label{eq:Egain}\end{equation}
where $\varepsilon$ is the dielectric permittivity of the surrounding medium. This energy gain is independent on the sign of the electric field, outwards or towards the surface as illustrated in Fig. \ref{Fig:F}.

\begin{figure}[b!]
\includegraphics[width=0.4\textwidth]{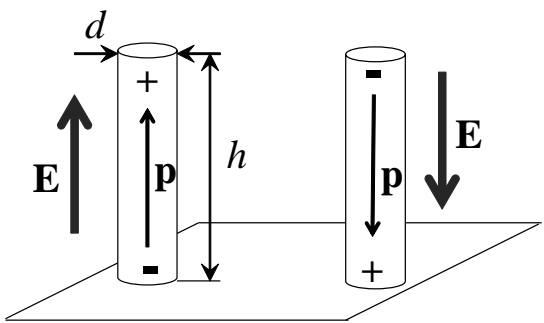}
\caption{Sketch of two whiskers of length $h$ and diameter $d$ on a metal surface with local electric fields ${\bf E}$ (of opposite directions) inducing the dipole moments ${\bf p}$. \label{Fig:whiskgeom}}
\end{figure}

$\alpha$ is a maximum in the longitudinal direction illustrated in Fig. \ref{Fig:whiskgeom} and is by approximately the factor of $(h/d)^2\gg 1$ greater than the particle volume $\pi (d/2)^2h$ that serves as a standard measure of polarizability in electrostatics. The mechanism of that enhancement can be understood as follows. Under an electric field $E$, a metallic needle will accumulate at its ends opposite charges of absolute values $q\sim Eh^2$ corresponding to the dipole moment $p\sim qH \sim Eh^3 \sim EV(h/d)^2$, where $V\sim hd^2$ is the particle volume.

More exactly (Ref. \onlinecite{landau1984}, p. 17) it is given by
\begin{equation}\alpha\approx \frac{h^3}{3\Lambda} \quad {\rm with} \quad \Lambda\equiv \ln(4h/d)-7/3.\label{eq:ncyl}\end{equation}
Note that the concept of energy gain in Eqs. (\ref{eq:Egain}) and (\ref{eq:ncyl}) has been used to describe the field induced nucleation of metal particles.\cite{karpov2007,karpov2008,karpov2008a,karpov2008b,nardone2009,nardone2012,nardone2012a,karpov2012,karpov2012a}
\bigskip

Side by side with the above energy gain, there is energy loss due to the whisker caused increase in the surface area, $W_A=\pi dh\sigma$  where $\sigma$ is the surface tension. The total change in free energy due to whisker formation is given by
\begin{equation}\label{eq:F}
F(h)=-\frac{h^3}{3\Lambda} \varepsilon E^2 +\pi dh\sigma.
\end{equation}
It is a maximum,
\begin{equation}\label{eq:nucbar}
\max{F(h)}= W\equiv \frac{2}{3}\pi\sigma d\sqrt{\frac{\pi\sigma\Lambda d}{\varepsilon E^2}}
\end{equation}
when
\begin{equation}
h=h_0\equiv\sqrt{\frac{\pi\sigma\Lambda d}{\varepsilon E^2}}.
\label{eq:nuclen}\end{equation}
 (Here we have treated a logarithmically weak dependence $\Lambda (h)$ as a constant.)
The barrier $W$ and its corresponding length $h_0$ have the same meaning as the nucleation barrier and radius in the classical nucleation theory. \cite{kaschiev2000} In particular, a whisker becomes stable and keeps growing when its length exceeds $h_0$, so it overcomes the energy barrier $W$.

Along the lines of standard nucleation theory, the above results introduce the characteristic nucleation time,
\begin{equation}\label{eq:nuctime}
\tau = \tau _0\exp\left(\frac{W}{kT}\right),
\end{equation}
upon which stable whisker with lengths $h>h_0$ can be observed. (We note parenthetically that the preexponential $\tau _0$ remains poorly determined in the framework of the existing classical nucleation theory, leading to many order of magnitude deviations from the data. Regardless, its often used values are ranging in the interval $\tau _0\sim 10^{-13}-10^{-8}$ s). As seen from Eq. (\ref{eq:nucbar}), the nucleation barrier $W$ is field dependent. Based on the consideration in Sec. \ref{sec:field} below, that field is a random variable; hence, nucleation times distributed in the exponentially broad interval. One other immediate prediction is that external fields (superimposed on the exiting random fields) can exponentially accelerate whisker nucleation.

The terminology of whisker nucleation in the existing literature \cite{cheng2011,nakai2009} was used as a qualitative statement discriminating between the stages of whisker conception and subsequent evolution. The approach in Eqs. (\ref{eq:F}) - (\ref{eq:nuclen}) provides a basis for the concept of whisker nucleation. This nucleation is triggered by the energy gain of metal whisker due to their polarization  in the surface electric field.

\section{The electric field distribution}\label{sec:field}
\begin{figure}[t!]
\includegraphics[width=0.40\textwidth]{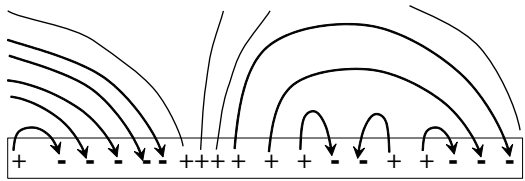}
\caption{Sketch of the the electric field lines in a system of randomly charged patches on a metal surface.\label{Fig:EE}}
\end{figure}

Sufficient electric fields above metal surface can arise from spatial variations of the work function. \cite{camp1991} The regions of different surface potential (patches) may be due to the polycrystallinity of the metal; the work function will vary between regions of specific grain orientations by typically a few tenths of a volt. Patch structure may also arise from the presence of adsorbed elements and compounds on the surface. The adsorption may be either chemical (oxidation or other reactions) or physical. Thus a surface that is initially electrically uniform may acquire surface structure upon exposure to air or other gases as illustrated in Fig. \ref{Fig:EE}.

\begin{figure}[b!]
\includegraphics[width=0.33\textwidth]{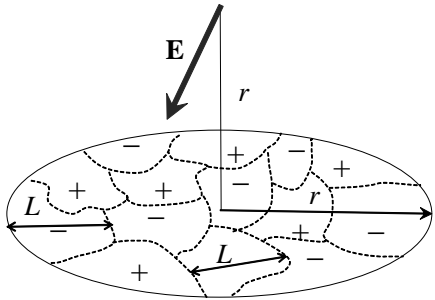}
\caption{Sketch of the patched area domain of radius $r$ where $+$ and $-$ represent positively and negatively charged patches (shown in dash) of characteristic linear dimension $L$ each. The fat arrow represents the random field vector at distance $r$ along the domain axis.\label{Fig:E}}
\end{figure}

The measurements reveal the typical work function fluctuations of $\sim 0.5$ eV induced by $\sim 10$ $\mu$m patches. \cite{camp1991} In general, the charged surface state concentration of $n\gtrsim 10^{12}$ cm$^{-2}$ not unusual for many materials \cite{sze} would correspond to the field strength of $E_0=4\pi en /\varepsilon\gtrsim 10^6$ V/cm where $e$ is the electron charge. This strong field orientations can be either up or downwards in Figs. \ref{Fig:EE} and \ref{Fig:E} extending over the characteristic distance $L$ above the surface.

As illustrated in Fig. \ref{Fig:E}, at distances $r\gg L$, the contributions of oppositely charged patches mostly cancel each other, and the field is due to an excess number $\Delta N$ of the patches of a certain sign close enough to the point of observation. Taking the latter at height $r$ above the surface, charged patches in a domain of radius $r$ beneath will generate more or less perpendicular random field. Therefore, $\Delta N\sim \sqrt{N}\sim r/L$, where $N\sim r^2/L^2$ is the average number of patches in the domain of radius $r$. As a result, one can estimate the absolute value of the projection of the field perpendicular to the surface,
\begin{equation}\label{eq:Eest}
E\sim \frac{\Delta Nne L^2}{r^2}\sim E_0\frac{L}{r}.
\end{equation}
The corresponding contribution to free energy is then estimated as [cf. Eqs. (\ref{eq:Egain}) and (\ref{eq:ncyl})]
\begin{equation}\label{eq:WE1}
W_E=- \frac{\varepsilon E_0^2hL^2}{3\Lambda}\quad {\rm when}\quad h\gg L .
\end{equation}

Far enough from the surface, the square of the field in Eq. (\ref{eq:Eest}) becomes very low. The alternative source is the background (thermal) electric field with time average
\begin{equation}
\langle E_T^2\rangle =4\pi\sigma _{SB}T^4/c\sim 20\quad {\rm V}^2{\rm cm}^{-2}.
\label{eq:Esq}\end{equation}
Here $\sigma _{SB}$ is the Stefan-Boltzmann constant, $c$ is the speed of light, and we chose the temperature $T\sim 300$ K. Comparing the results in Eqs. (\ref{eq:Eest}) and (\ref{eq:Esq}) yields the overplay distance
\begin{equation}\label{eq:rc}
r_c\sim L\frac{E_0}{E_T}.
\end{equation}
For the above mentioned numerical values, this length ($\sim 10$ cm) is far beyond the whiskers length domain. However it could shrink down to that domain for the case of very high temperatures or low surface state densities.

The region of $r\gg L$ is irrelevant for nucleation events (we will see that $h_0\ll L$). However, it can be important for the growth stage of whisker formation when their lengths $h$ exceed $L$. Also, for a hypothetical case of rather small patches, say, $L\sim 100$ nm, the length $r_c\sim 0.1$ mm would fall in the domain of the observed whisker dimensions.

Overall, we conclude that the strong field region extends up to the patch length $L$ above metal surface and that the field in that region is more or less perpendicular to the surface. At distances $r\gg L$ but $r\ll r_c$, the field is on average reciprocal of $r$ and its directions are more random. Finally for very large $r\gg r_c$, the field strength is determined by the background radiation that is polarized parallel to the surface. \cite{rytov}
\section{Whisker growth}
A nucleated whisker will grow thereby further decreasing its free energy $F$. In the Fokker-Planck approximation, \cite{landau2008} its average length increases in time $t$ according to
\begin{equation}\label{eq:FP}
\frac{dh}{dt}=-b\frac{dF}{dh}
\end{equation}
where $b$ is the generalized mobility related, via Einstein formula, to the diffusion coefficient in the whisker size space,
\begin{equation}\label{eq:mob}
b=\frac{D}{kT}\quad {\rm where} \quad D=D_0\exp\left(-\frac{E_d}{kT}\right).
\end{equation}
Here $E_d$ is the diffusion activation energy, and $D_0$ remains an unknown parameter. Base on the purely dimensional argument, it  must be not very different from the preexponential of the diffusion responsible for whisker growth.

The diameter will increase as well in the course of whisker growth. However, the analysis of evolution of both dimensions is more complex and falls beyond the scope of this exploratory work; here we limit ourselves to the discussion of $h(t)$.

Integrating Eq. (\ref{eq:FP}) with $F=-W_E$ (i. e. neglecting surface energy far enough from the nucleation barrier) and $W_E$ from Eqs. (\ref{eq:Egain}) and (\ref{eq:WE1}) yields
\begin{equation}\label{eq:t0}
h=\frac{h_0}{1-t/t_0}\quad t_0\equiv \frac{3\Lambda}{b\varepsilon E_0^2h_0}\quad {\rm when}\quad h\ll L,
\end{equation}
and
\begin{equation}\label{eq:tL}
h=L\frac{t}{t_L}\quad t_L\equiv \frac{3\Lambda}{b\varepsilon E_0^2L}\quad {\rm when}\quad r_c\gg h\gg L.
\end{equation}
Finally, in  the region of yet larger lengths, $h\gg r_c$, the whiskers will grow uniformly, as predicted by Eq. (\ref{eq:tL}) where $E_0^2$ is replaced with $\langle E_T^2\rangle$.

A comment is in order with regards to the above description of whisker growth that is essentially one-dimensional, along the coordinate perpendicular to metal surface. In reality, the patch induced field can have significant lateral components forcing whisker development (along the field lines) in the lateral directions. The degree of the corresponding winding will be determined by the competition between the electrostatic energy gain and energy loss due to the filament bending. A more exact description of this fascinating phenomena of filament growth in a labyrinth of  random electric fields, while can be developed, falls far beyond the limited scope of this work.

The latter comment should extend over the region of very long whiskers, $h\gg r_c$, where whisker growth is dominated by the field that is strictly parallel to the metal surface. One can expect then that having reached the interplay length $h\sim r_c$, the whisker shape should deviate from rectilinear most noticeable, showing spiraling or random walking configurations.

\section{Numerical estimates and discussion}
We shall use the above mentioned parameter values: $E_0=3\times 10^5$ V/cm, $L=10$ $\mu$m, $T=300$ K, and $\varepsilon =1$. Also, we shall use \cite{saka1988} $\sigma =30$ erg/cm$^2$, the diffusion coefficient for tin \cite{woodrow2006} $D\sim 10^{-18}$ cm$^2$s$^{-1}$, and approximate $\Lambda =1$ (consistent with these values). Also, following multiple examples of applications of the field induced nucleation, \cite{karpov2007,karpov2008,karpov2008a,karpov2008b,nardone2009,nardone2012,nardone2012a,karpov2012,karpov2012a} we consider $d$ to be a minimum diameter still consistent with the concept of metal wire, that is of the order of $d=0.5$ nm.

With the above parameters in mind, Eqs. (\ref{eq:nucbar}) and (\ref{eq:nuclen}) yield $W=0.6$ eV and $h=15$ nm, the latter corresponding to high aspect ratio $h/d=30$ consistent with the concept of needle shaped whisker nucleation. The latter $W$ is relatively small against the scale of many other nucleation barriers $W\sim 1-3$ eV known for various phase transformations. Based on these estimates, the whisker nucleation time is expected to be relatively very short, in the subsecond range.

The characteristic times describing whisker growth, from Eqs. (\ref{eq:t0}) and (\ref{eq:tL}) are estimated as $t_0\sim 10^4$ s, and $t_L\sim 10^2$ s. We should remember that $t_L$ can be severely underestimated by neglecting the 3D nature of electric field fluctuations and the corresponding winding of whiskers. Finally, the latest stage of whiskers growth upon reaching the critical length $r_c$ is described by the characteristic time that is by the factor $E_0^2/\langle E_T^2\rangle\sim 10^5$ longer, i. e. $t_T\sim 10^7$ s.

With the above reservation in mind, these estimates result in the following scenario: (i) stage 1 -- whiskers nucleate in a sub-second to many days time interval reflecting fluctuations in their nucleation barriers related to the local field fluctuations; (ii) stage 2 -- they grow up to the patch linear dimension, say 10 $\mu$m, (more or less perpendicular to the surface) during much longer time $t_0\sim 10^4$ s that can be experimentally identified with an incubation time, and, again, fluctuates due to the local field fluctuations; (iii) they grow above 10 $\mu$m in a relatively short time $t_L$ (that can be much longer depending on the whiskers bending parameters); (iv) if the system parameters are such that feeding by thermal radiation is possible, then further whisker growth and entangling takes place on the scale of months.

Because a significant effort was spent to understand whiskers in terms of mechanical stresses, recrystallization, dislocations, etc., it should be noted that the present theory does not rule out these factors. Furthermore, they can be a part of the picture presented, resulting, for example, in local spots of unfavorable energy configurations capable of relaxing through the mechanism of field induced nucleation, and/or local spots of stress induced electricity. In particular, it was shown that mechanical stresses generally strongly affect metal surface corrosion rates, which observation points at the stress induced change of the surface electric potential. These changes can be caused by dislocations, \cite{yin2007} stress induced spots of different structure phases, \cite{namahoot2004} or general electric deformation coupling \cite{kolodii2000} in combination with stress induced buckling. \cite{balakrisnan2003,wertheim1991} Local charges due to stress induced oxide cracking on or ion trapping under the whisker growing metal layer (say, Sn on Cu substrate) are conceivable sources of the above considered surface electric fields as well.

\section{Conclusions}
This manuscript attempts a novel whisker theory based on the electrostatic driving force -- never considered before in connection with whiskers. From a broader perspective, it is an extension of the recently developed theory of field induced nucleation \cite{karpov2007,karpov2008,karpov2008a,karpov2008b,nardone2009,nardone2012,nardone2012a,karpov2012,karpov2012a} over a fascinating science of whiskers (that remained largely overlooked by physicists). One other broad observation is that it presents the first whisker theory yielding simple analytical results that are at least semi-quantitatively consistent with the observations.

More specifically, the above theory answers the following questions.\\
1) Why whiskers are metallic: high (metallic) electric polarizability is required for sufficient energy gain due to whisker formation.\\
2) Why whiskers grow predominantly perpendicular to the surface: such are the dominating directions of the surface electric field.\\
3) Why whisker parameters are broadly statistically distributed: this reflects fluctuations in metal surface fields due to various imperfections.\\
4) Why some metals are more prone to develop whiskers: because they can easier adsorb ions or grow oxides forming charged patches on their surfaces.\\
5) Why contaminating these metal surfaces with various ions triggers whisker growth, while treatments with DI water or other appropriate cleaning suppress whisker growth: again, adding or removing sources of surface electric fields.\\
6) Why external electric biasing significantly accelerates whisker growth: the electric field increases nucleation and growth rates.\\
7) Why some minute additives can significantly affect whiskers: through changes in surface tension and surface electric field distributions.\\
8) Why the grain size is not a significant whisker factor: it is not strongly related to the surface electric field.\\
9) Why the characteristic whisker evolution rates are such as they are: the important physical parameters are explicitly included in the above equations for the characteristic times. In particular, the observed long incubation period ($t_0$) followed by a relatively fast linear growth ($t_L$) with saturation ($t_T$) is explained.\\
10) Correlation between whiskers and (i) grains whose orientation is different from the major
orientation of the tin film, (ii) dislocations and dislocation loops, and (iii) mechanical stresses capable of surface buckling: all these factors related to local surface charges and their whisker provoking electric fields.

A set of here predicted dependencies of nucleation and growth kinetics vs. electric field, temperature, and controlled contamination could be verified experimentally. It should be noted however that this work presents rather a sketch of theory in its infancy, pointing at important factors and providing rough estimates, yet not enough developed to quantitatively describe whisker evolution and statistics in a random electric field. Further effort is called upon to develop this approach.

I would like to finish this manuscript by pointing at a prediction of active means affecting whisker conception and allowing well controlled growth of metal nanowires of desirable parameters on any metal surface. This can be achieved by creating the surface plasmon polariton excitations known to induce anomalously strong electric fields above metal surfaces. \cite{maier2007} Along these lines, electric field sufficient for whisker production can be achieved not only for selected materials (Sn, Zn, Cd, some others), but for any metal of interest. This prediction appears verifiable with the commonly available equipment.

\section*{Acknowledgement}

Multiple discussions with D. Shvydka, M. I. Karpov (Goldfeld), A. V. Karpov, A. V. Subashiev, A. B. Pevtsov, and D. A. Parshin are highly appreciated.

\end{document}